\documentclass[journal]{IEEEtran}

\pdfoutput=1

\usepackage{cite}
\usepackage{url}

\usepackage{enumitem}

\ifCLASSINFOpdf
    \usepackage[pdftex]{graphicx}
\else
    \usepackage[dvips]{graphicx}
\fi

\usepackage{amsmath}

\usepackage{makecell}

\begin{document}

\title{Reflectionless Filters for Generalized Transmission Functions}

\author{Augusto~Guilabert, Matthew~A.~Morgan,~\IEEEmembership{Senior Member,~IEEE}, and Tod A. Boyd%
\thanks{Manuscript received ???}
\thanks{A. Guilabert, now retired, was previously development engineer in the telecom industry with Standard El\'ectrica S.A.(ITT Corp.), Alcatel N.V. and 3Bymesa S.A. (e-mail: a.guilabert@3bymesa.com)}
\thanks{M. Morgan and T. Boyd are with the Central Development Laboratory, National Radio Astronomy Observatory, Charlottesville,
VA, 22903 USA (e-mail: matt.morgan@nrao.edu). The National Radio Astronomy Observatory is a facility of the National Science Foundation operated under cooperative agreement by Associated Universities, Inc.}}

\markboth{IEEE Transactions on Circuits and Systems-I,~Vol.~xx, No.~x, xxx~2019}%
{Guilabert \MakeLowercase{\textit{et al.}}: Reflectionless Filters for Generalized Transmission Functions}

\IEEEpubid{0000--0000/00\$00.00~\copyright~2019 IEEE}


\maketitle

\begin{abstract}
Single-ended circuit topologies, and a theorem for the development thereof, are presented with which one may realize constant-resistance (or reflectionless) filters, having ideally zero reflection coefficient at all frequencies and from all ports, suitable for elliptic and pseudo-elliptic filter responses. The proposed theorem produces topologies of a type known as the coupled-ladder, which has been previously studied for only polynomial responses (e.g. Butterworth, Chebyshev, etc.). A comparison between these topologies and another classical approach known as the economy bridge reveals that those proposed here have a number of theoretical and practical advantages. The theory is tested by the construction of a sixth-order, low-pass reflectionless filter exhibiting a pseudo-elliptic frequency response. Measured results are in excellent agreement with theory, and show return loss better than 20 dB throughout the pass-band, the transition-band, and up to two octaves into the stop-band.
\end{abstract}


\section{Introduction}

\IEEEPARstart{C}{onstant} resistance, or \emph{reflectionless}, filters make use of lattice networks where the arms $Z_A$ and $Z_B$ are conventional dual reflective filters (to be named \emph{reference} filters). An unbalanced equivalent, presented by Cauer in \cite{cauer} uses a single \emph{economy bridge} with only two arms, each comprising one of the dual reference filters. The characteristic function of these filters must be the reciprocal of that desired for the transmission path of the global network. Another method to obtain a practical unbalanced structure, referred to herein as a \emph{coupled-ladder}, has been developed in \cite{morgan_theoretical, morgan_structures, morgan_artech, morgan_ladder} and  makes use of the progressive construction of the reflectionless network from even-/odd-mode analysis operating on the ladder topology itself. Applications of this coupled-ladder synthesis operating with polynomial reference filters have been studied in \cite{morgan_ladder}. The more general case, e.g. filters having equiripple in both the pass-band and stop-band as provided by Cauer elliptic functions, or of any arbitrary suitable configuration of poles and reflection zeroes on the $j\omega$ axis, will be addressed here.

First, a review of the conceptual scheme underlying coupled-ladder synthesis, set forth by an equivalence theorem, is given in Section~\ref{sec:theorem}. The theorem is this time more easily approached by an algebraic method instead of the traditionally used even-/odd-mode analysis. A comparison between the economy bridge and coupled ladder synthesis results will be given in Section~\ref{sec:eb_comparison}. In Section~\ref{sec:elliptic}, specific issues concerning the realization of elliptic or quasi-elliptic cases will be addressed under the guidance of the presented theorem. Finally, Section~\ref{sec:practical_example} will describe a practical realization in hardware along with the measured results.

A well-known predecessor of the coupled-ladder method is the lattice reduction technique used to obtain an unbalanced equivalent of a given lattice \cite{weinberg1951new}. The relationship between both methods will be commented upon in Appendix~\ref{app:lattice_reduction}.

\section{A Realization Theorem for Lattice Networks with Ladder Arms}\label{sec:theorem}

Here, we present a theorem for the construction of a single-ended circuit equivalent in performance to a balanced lattice network with ladder arms, as an alternative to Cauer's economy bridge and having certain practical advantages which will be described in Section~\ref{sec:eb_comparison}. For the derivation, the approach this time will consist of an algebraic method, instead of even-/odd-mode analysis, but the resulting family of networks will be the same as the one studied in \cite{morgan_theoretical, morgan_structures, morgan_artech, morgan_ladder}.

\subsection{Preliminary Definitions}

\IEEEpubidadjcol

\begin{itemize}
    \item Any two ladders A and B may be driven to an isomorphic form, where both have an identical number of branches grouped into \emph{cells} by pairs, one from each ladder, where the two branches paired in each cell have both to be series or shunt in the corresponding ladder, and in the same relative position. Cells are then characterized as being of the \emph{series} or \emph{shunt} type. Zero impedance or zero admittance branches are admitted  in series or shunt cells, respectively. It should be noted that the isomorphic form is not unique, as different associations are possible. See the example in Fig.~\ref{fig:isomorphism}.
    \begin{figure}[!t]
        \centering
        \includegraphics{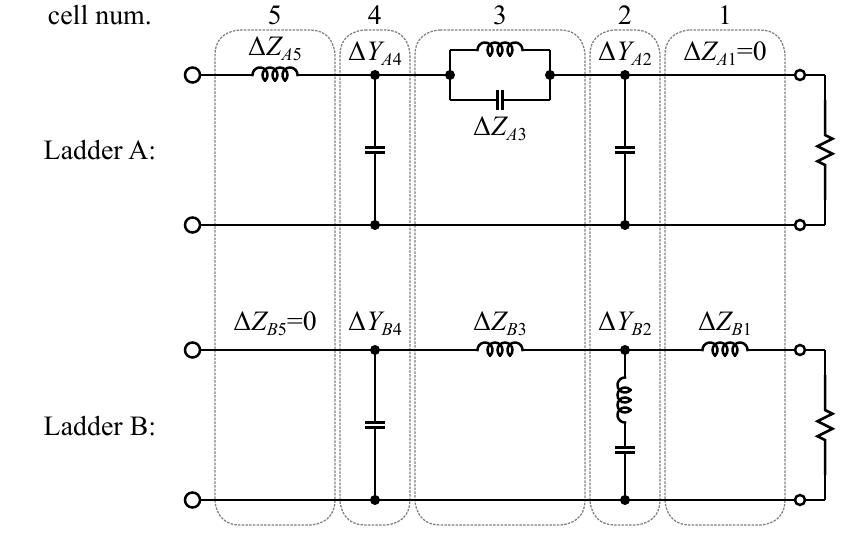}
        \caption{Establishing ladder isomorphism. Each grey box encircles a \emph{cell}.}
        \label{fig:isomorphism}
    \end{figure}
    \item $\Delta Z_{Ai}$ and $\Delta Z_{Bi}$ (or $\Delta Y_{Ai}=1/\Delta Z_{Ai}$ and $\Delta Y_{Bi}=1/\Delta Z_{Bi}$) are the impedances (admittances) of  cell $i$ in ladders A and B, respectively.
    \item Cells shall be indexed consecutively, starting at the termination, independent of whether they are the series or shunt type; they do not need to be alternating, though this is the usually the case. 
    \item \emph{Elementary lattice} $J$ is a network equivalent to the lattice constructed with arms corresponding to cell $J$.
    \item The two normalized termination resistances can be interpreted as a termination lattice with a $z$- or $y$-parameter matrix given by the 2x2 identity matrix, $\mathbf{I}$.
    \item The input impedance of reference ladder A may be written as a finite continued fraction, whose exact form will depend on whether the first and last elements are series or shunt.
    In case (a), where the first and last cells are both series,
    \begin{equation}
        Z_A = \Delta Z_{An}+\dfrac{1}{\Delta Y_{A(n-1)}+\dfrac{1}{\dfrac{\ddots}{\Delta Y_{A2}+\dfrac{1}{\Delta Z_{A1}+1}}}}
    \end{equation}
    Similarly, in case (b), where both are shunt,
    \begin{equation}
        Y_A = \Delta Y_{An}+\dfrac{1}{\Delta Z_{A(n-1)}+\dfrac{1}{\dfrac{\ddots}{\Delta Z_{A2}+\dfrac{1}{\Delta Y_{A1}+1}}}}
    \end{equation}
    and so on for the remaining two cases (c) and (d), where the first/last elements are series/shunt, and shunt/series, respectively. Similar expressions hold for $Z_B$ and $Y_B$.
    \item Let $\mathbf{Z}_{NW}$ be the $z$-parameter matrix of the global lattice network NW comprising the complete ladders A and B, having impedances $Z_A$ and $Z_B$, as the main arms,
    \begin{equation}\label{eq:ZNW}
        \mathbf{Z}_{NW} = \frac{1}{2}\begin{pmatrix}
            Z_B+Z_A & Z_B-Z_A\\
            Z_B-Z_A & Z_B+Z_A\end{pmatrix}
    \end{equation}
\end{itemize}

\subsection{Equivalence Theorem}\label{sec:equivalence_theorem}

The theorem expresses the validity of the following alternative development of $\mathbf{Z}_{NW}$. For case (a) above (first and last cells both series),
\begin{equation}\label{eq:NW_zparams}
    \mathbf{Z}_{NW} = \Delta\mathbf{Z}_n+\dfrac{\mathbf{I}}{\Delta\mathbf{Y}_{n-1}+\dfrac{\mathbf{I}}{\dfrac{\ddots}{\Delta\mathbf{Y}_2+\dfrac{\mathbf{I}}{\Delta\mathbf{Z}_1+\mathbf{I}}}}}
\end{equation}
where $\mathbf{I}$ is again the 2x2 identity matrix and the fraction notation shall be understood to mean the matrix inverse.
For case (b), where the first and last elements are both shunt,
\begin{equation}\label{eq:NW_yparams}
    \mathbf{Y}_{NW} = \Delta\mathbf{Y}_n+\dfrac{\mathbf{I}}{\Delta\mathbf{Z}_{n-1}+\dfrac{\mathbf{I}}{\dfrac{\ddots}{\Delta\mathbf{Z}_2+\dfrac{\mathbf{I}}{\Delta\mathbf{Y}_1+\mathbf{I}}}}}
\end{equation}
and similarly for cases (c) and (d). In the above expressions, $\Delta\mathbf{Z}_j$ and $\Delta\mathbf{Y}_k$ are the $z$- and $y$-parameter matrices of the elementary lattices corresponding to cell $j$ and $k$. That is,
\begin{subequations}\label{eq:elementary_zparams}
    \begin{equation}
        \Delta\mathbf{Z}_j = \frac{1}{2}\begin{pmatrix}
        \Delta Z_{Bj}+\Delta Z_{Aj} & \Delta Z_{Bj}-\Delta Z_{Aj}\\
        \Delta Z_{Bj}-\Delta Z_{Aj} & \Delta Z_{Bj}+\Delta Z_{Aj}
        \end{pmatrix}
    \end{equation}
    \begin{equation}
        \Delta\mathbf{Y}_k = \frac{1}{2}\begin{pmatrix}
        \Delta Y_{Bk}+\Delta Y_{Ak} & \Delta Y_{Bk}-\Delta Y_{Ak}\\
        \Delta Y_{Bk}-\Delta Y_{Ak} & \Delta Y_{Bk}+\Delta Y_{Ak}
        \end{pmatrix}
    \end{equation}
\end{subequations}
Equations \eqref{eq:NW_zparams}-\eqref{eq:elementary_zparams} show that that a two-port equivalent to the lattice-network NW can be constructed alternatively as a ladder connection of elementary lattices corresponding to the cells established in the previous isomorphic definition, with the same order and type of connection. The proof of the theorem is straight-forward, and is summarized in Appendix~\ref{app:demonstration}.

Fig.~\ref{fig:lattice_equivalents}
\begin{figure}[!t]
    \centering
    \includegraphics{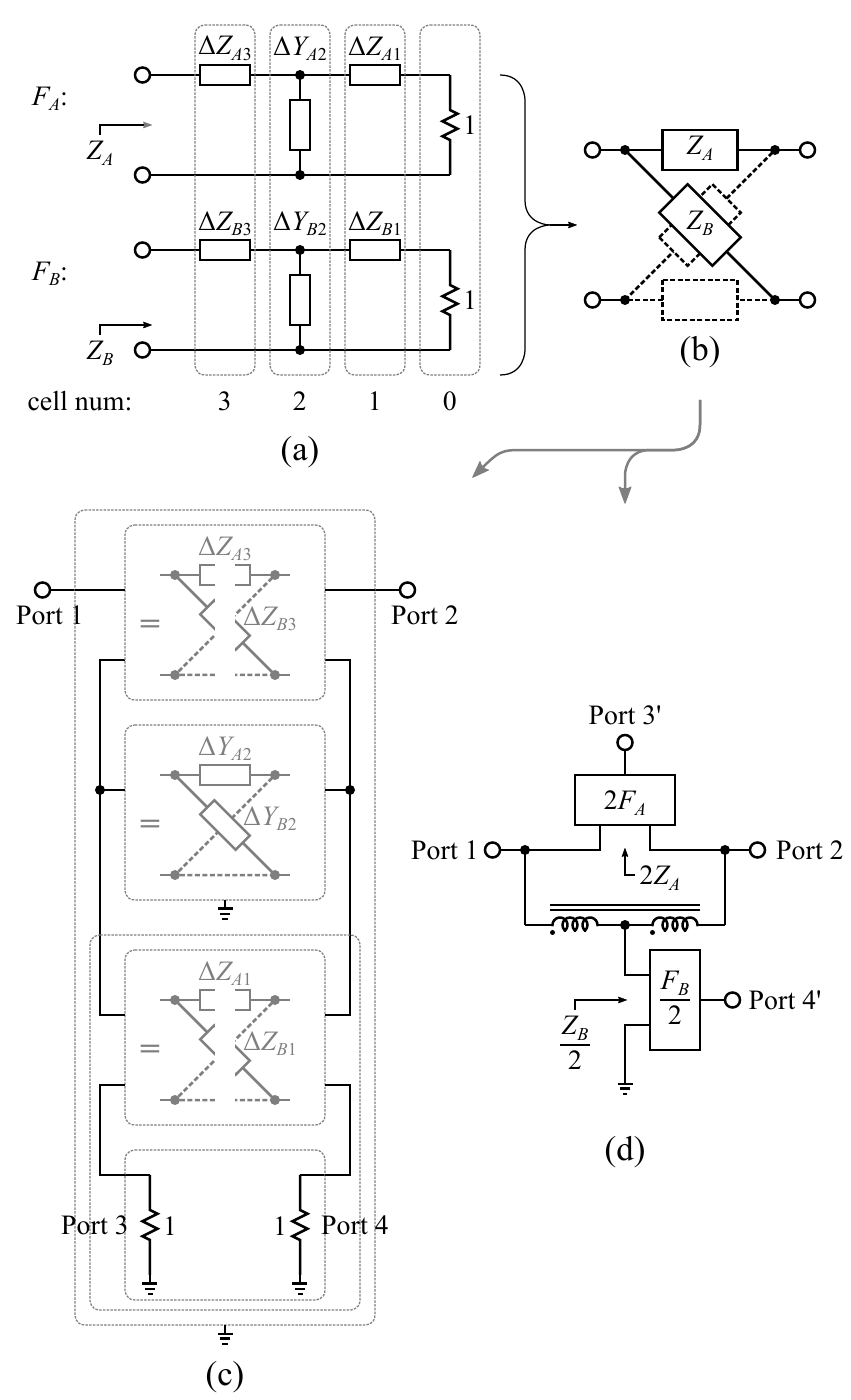}
    \caption{Lattice equivalent constructions when arms are ladder networks. (a) Decomposition of ladder arm immittances into cells. (b) Lattice network comprising four complete ladders. (c) Coupled-ladder equivalent construction from elementary lattices. All connections, frames, and elements in grey express that their meaning is conceptual, or ``equivalent to''. (d) Cauer's economy bridge equivalent using a single ideal transformer. Ports 3' and 4' are the termination resistors on ladders $2Z_A$ and $Z_B/2$, respectively.}
    \label{fig:lattice_equivalents}
\end{figure}
illustrates the theorem in graphical form. Fig.~\ref{fig:lattice_equivalents}(a) shows the reference filter ladders with impedance $Z_A$ and $Z_B$ broken down into series and shunt cells. For the purpose of this illustration, a simple case of three cells has been drawn, with the cell number 1, adjacent to the terminations, being series type. Fig.~\ref{fig:lattice_equivalents}(b) shows the balanced lattice network we wish to draw an equivalent to. Fig.~\ref{fig:lattice_equivalents}(c) illustrates, conceptually, the construction of a coupled ladder of elementary lattice \emph{equivalents} from bottom to top in series and shunt fashion in accordance with the cells of the reference ladders. All connections, frames, and elements in grey indicate that the given schematics are conceptual, or \emph{equivalent to}. For instance, the series connection of two-ports may be stacked vertically as drawn, or by any other convenient and known method of network combination that leads to the addition of $Z$ parameters. The necessary condition for lateral current continuity of the lattice equivalent is symbolically represented in the series cases by a drawing cut. The resulting partial network after each step is framed again in grey.

A summary of elementary lattice equivalent networks which can be used in this construction process for series and shunt cells will be given in Tables~\ref{tab:series} and \ref{tab:shunt} of Section~\ref{sec:elliptic}, respectively. Typical application cases are included. The incremental ladder immittances in these tables are labelled $Z_a$, $Y_a$, etc.

Briefly put, the theorem states that a lattice of two ladders is equivalent to a ladder of elementary lattices. Further implications in more mathematical terms include:
\begin{itemize}
\item In the scope of two-port equivalence, successive application of ladder and lattice ``two-port construction'' operators to a set of indexed, paired, two-poles (with genders series and shunt) obey the commutative property.
\item For a lattice network, the availability of the continued fraction expansions of arm immittances leads directly to the same kind of realizable expansion of the global lattice $Z$ (resp. $Y$) matrix. The corresponding transformation preserves ladder branch entities.  
\end{itemize}
In the praxis, it allows the direct breakdown of a lattice with complex arms (e.g. ladder filters) into simple interconnected lattices or lattice-equivalents. 

The classical  alternative to the lattice network is the Cauer economy bridge, shown in Fig.~\ref{fig:lattice_equivalents}(d). Although easier to draw, conceptually, the simplicity of this diagram is somewhat deceptive considering the number of elements that may be contained in $2F_A$ and $F_B/2$ ladders. Furthermore, there are number of drawbacks, both fundamental and practical, with this approach. A thorough comparison of the coupled-ladder and economy bridge topologies will be given in Section~\ref{sec:eb_comparison}.

\begin{figure}[!t]
    \centering
    \includegraphics{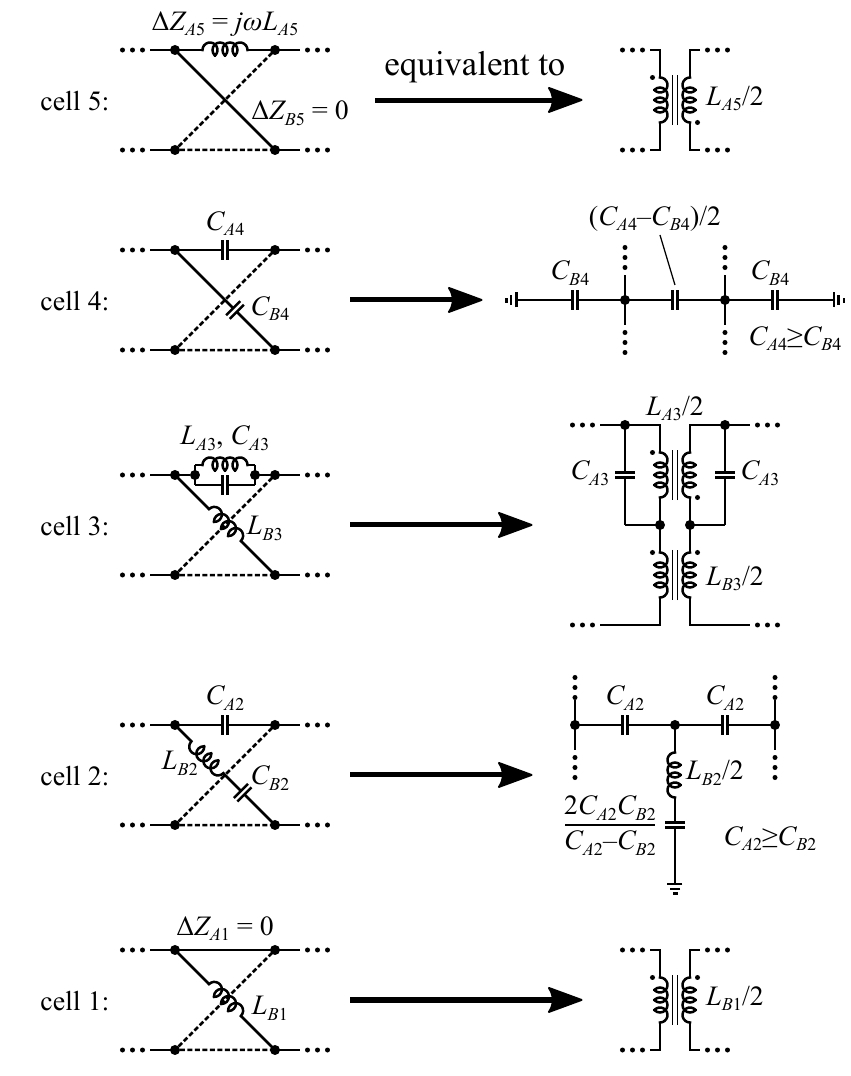}\\
    (a)\\
    \includegraphics{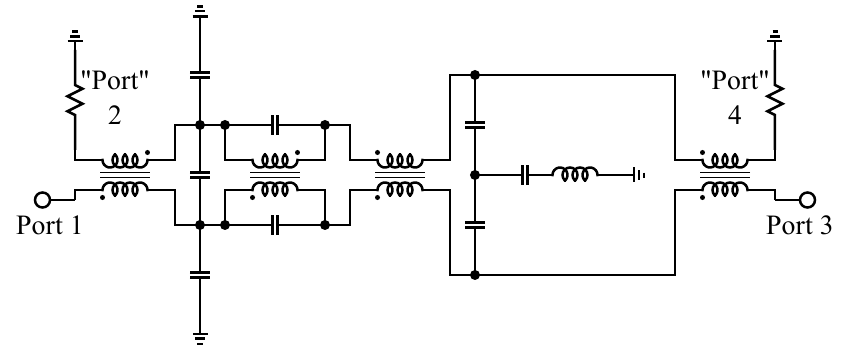}\\
    (b)
    \caption{A possible coupled-ladder realization using the ladders of Fig.~\ref{fig:isomorphism}. (a) Equivalent networks for cell elementary lattices. (b) Fully-assembled network (rotated for easy viewing).}
    \label{fig:possible_realization}
\end{figure}

Some important features to note of the theorem we have presented:
\begin{enumerate}
    \item The equivalence is valid in general; no requirement has been made on duality, but when ladders $F_A$ and $F_B$ do behave as dual two-ports, reflectionless networks result. Black-box duality --- that is, dual electrical \emph{behavior} as opposed to topology --- is sufficient in this case. 
    \item It is a simple matter to show that each network-half of a constructed network reproduces the structure of reference ladder $F_B$ in the even-mode, and that of ladder $F_A$ in the odd-mode. The global network satisfies, then, the main design criterion used in \cite{morgan_theoretical, morgan_structures, morgan_artech, morgan_ladder}, constituting an extension of the network family studied therein.
    \item In practice, having cells with proportional impedances --- that is, $\Delta Z_{Bi} = \alpha\Delta Z_{Ai}$ where $\alpha$ is real and positive --- may lead to more simple and realizable associated elementary two-ports, without need of magnetic coupling. This case corresponds to realizable tee- or pi-equivalent networks. Potentially harmful side effects of certain coupling-free realizations (and some with coupling as well) will be discussed in Section~\ref{sec:elliptic}.
    \item The rules of the known lattice-reduction technique (summarized in Appendix~\ref{app:lattice_reduction}) can be considered as particular implementation cases of the presented concept.
    \item The theorem has general coverage in that it is equally applicable to non-proportional $Z$ ($Y$) cell immittances, and the corresponding elementary lattice equivalents are realizable, even without coupling in some cases (e.g.: Fig.~\ref{fig:possible_realization}).
\end{enumerate}

Once the reference filter isomorphism and elementary lattices are defined, the application of well-known combination rules for the safe series and parallel connection of those lattices, together with their practical transformation according to Tables~\ref{tab:series} and \ref{tab:shunt}, will generate the network. Typically, some of the stages will require coupling while others do not.

As an example, consider the application of this process to the reference ladders in Fig.~\ref{fig:isomorphism}. The elementary lattices for each cell along with their equivalent networks are shown in Fig.~\ref{fig:possible_realization}(a)
and the final resulting coupled-ladder network is shown in Fig.~\ref{fig:possible_realization}(b). Other realizations are also possible, using elementary lattice equivalents given in Tables~\ref{tab:series} and \ref{tab:shunt} of Section~\ref{sec:elliptic}.

\section{Comparison with Cauer's Economy Bridge}\label{sec:eb_comparison}

An alternative, single-ended network configuration is Cauer's economy bridge, shown earlier in Fig.~\ref{fig:lattice_equivalents}(d). The theorem in Section~\ref{sec:theorem} establishes the equivalence of the coupled-ladder to the economy bridge, where ports 1 and 2 are the input and output ports; their transmission from port 1 to port 2 is the same. Nothing is stated, however, about the transmission from port 1 to port 3 (or port 1 to port 4) in the coupled-ladder, or from port 1 to port 3' (port 1 to port 4') in the economy bridge. As we shall see, the two networks behave quite differently in this respect.

In the economy bridge structure, if ladders $Z_A$ and $Z_B$ are dual reflective filter structures, which themselves realize a certain transmission function e.g. high-pass, then power is equally divided into both (high-pass) ladders, and power arriving to the output ports 3' and 4' will have additional 3 dB flat-loss. This is a clear limitation of economy bridge structures. Further, ports 3' and 4' do not exhibit reflectionless behavior. In fact, the full, four-port scattering parameter matrix for the economy bridge with dual arms is
\begin{equation}\label{eq:SEB}
    \mathbf{S}_{EB} = \begin{pmatrix}
    0 & \rho_d & \tfrac{\tau}{\sqrt{2}} & \tfrac{\tau}{\sqrt{2}}\\
    \rho_d & 0 & -\tfrac{\tau}{\sqrt{2}} & \tfrac{\tau}{\sqrt{2}}\\
    \tfrac{\tau}{\sqrt{2}} & -\tfrac{\tau}{\sqrt{2}} & \rho_r & 0\\
    \tfrac{\tau}{\sqrt{2}} & \tfrac{\tau}{\sqrt{2}} & 0 & -\rho_r
    \end{pmatrix}
\end{equation}
where $\tau$, $\rho_d$, and $\rho_r$ are specific functions of the even-mode reference filter. Using the notation established in \cite{christian} and \cite{Temes}, where $E(s)$, $F(s)$, and $P(s)$ are the transfer polynomials of the even-mode reflective filter, $\tau(s)$ is the inverse of the transducer function $H(s)$, or
\begin{equation}
    \tau(s) = \frac{P(s)}{E(s)} = \frac{1}{H(s)}
\end{equation}
Further, the input reflection coefficient of the reference filter is
\begin{equation}
    \rho_d = \frac{F(s)}{E(s)}
\end{equation}
and the output reflection coefficient of the reference filter (see \cite{Temes}, p. 82) is
\begin{equation}
    \rho_r = \frac{\mp F(-s)}{E(s)}
\end{equation}
where the `$-$' above is used if $P(s)$ is even, and `$+$' if $P(s)$ is odd.

In contrast, for case of the coupled ladder, also constructed from dual reference filters, both transmission paths 1\nobreakdash-2 and 1\nobreakdash-3 are candidates to be exploited, since both ideally realize 0\nobreakdash-dB loss in pass-band. The scattering parameter matrix in this case is simply
\begin{equation}\label{eq:SCL}
    \mathbf{S}_{CL} = \begin{pmatrix}
        0 & \rho_d & \tau & 0\\
        \rho_d & 0 & 0 & \tau\\
        \tau & 0 & 0 & \rho_r\\
        0 & \tau & \rho_r & 0
    \end{pmatrix}
\end{equation}
which is clearly distinct from \eqref{eq:SEB}. If the coupled-ladder 1\nobreakdash-2 signal path is chosen, the same high-pass reference filters of the economy bridge design can be used, for a desired low-pass transmission function. If, however, the coupled-ladder 1\nobreakdash-3 path is chosen, the reference filters must have exactly the same transmission function, $\tau(s)$, as that which is desired for the overall reflectionless filter. As previously shown \cite{morgan_structures}, and explicit in \eqref{eq:SCL}, the coupled-ladder structure exhibits directional behavior, wherein no power is transmitted from port 1 to 4 (they are decoupled, and also between 2 and 3), and the desired transmission function appears from 1 to 3, with no flat loss and constant resistance at all ports (not the case in economy bridge). The derived coupled-ladder structure belongs to a wider class of reflectionless reciprocal and lossless 4-ports, named \emph{biconjugate} networks. See \cite{belevitch1958} for a clever theoretical presentation.

One could argue that the comparison of the two structures should be limited to their 0-dB transmission paths only (1\nobreakdash-2 for the economy bridge, and 1\nobreakdash-2 or 1\nobreakdash-3 for the coupled ladder). The 1\nobreakdash-2 path, however, seems to have inherent practical disadvantages for both structures. A desired low-pass behavior between 1\nobreakdash-2 would require two dual reflective reference high-pass designs. The high-pass reflective structures, however, have unavoidable matching degradation at high frequencies due to strays. That inevitably results in degraded suppression in the stop-band at high-frequency \cite{morgan_ladder}.

\begin{figure}[!t]
    \centering
    \includegraphics{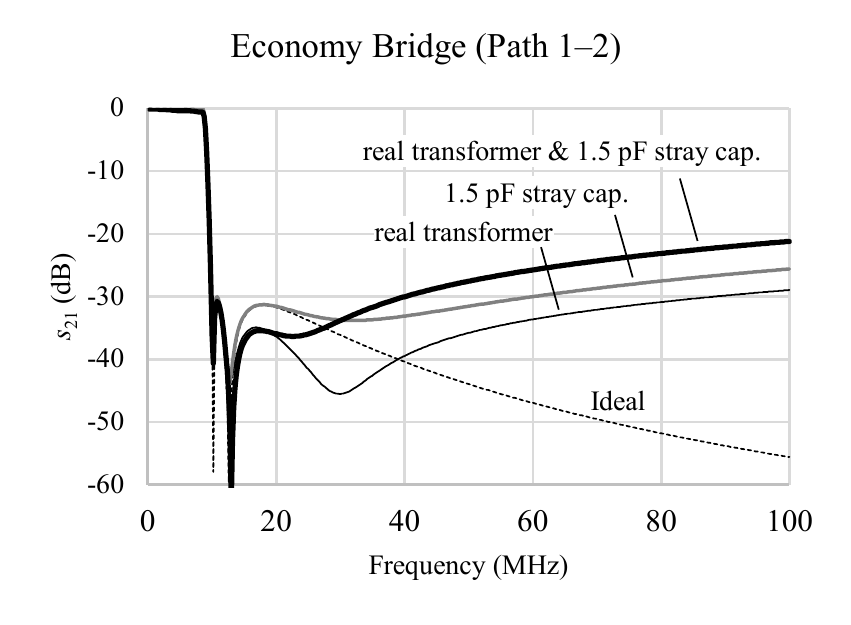}\\
    (a)\\
    \includegraphics{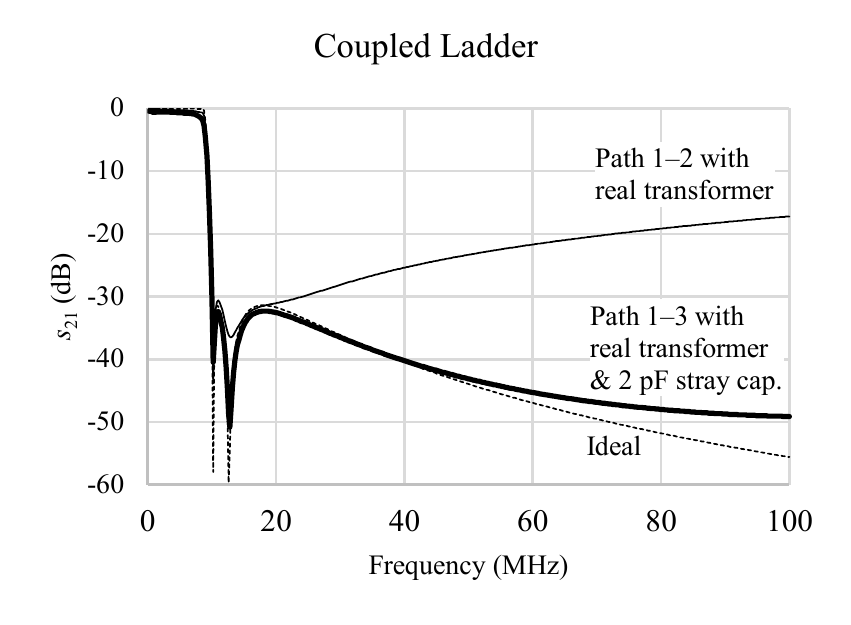}\\
    (b)
    \caption{Comparison of (a) economy bridge path 1-2 and (b) coupled-ladder (both cases, 1-2 and 1-3) with and without critical parasitics.}
    \label{fig:transmission_response_comparison}
\end{figure}

In fact, stop-band rejection is totally reliant on bridge symmetry in the case of the economy bridge structure. This seems to indicate a theoretical preference for the coupled-ladder 1\nobreakdash-3 path, which is illustrated by the practical simulation in Fig.~\ref{fig:transmission_response_comparison}.
This figure compares the 1\nobreakdash-2 transmission response of an economy bridge structure realizing a sixth-order antimetric low-pass filter (having two finite poles), with the 1\nobreakdash-2 and 1\nobreakdash-3 transmission responses of coupled-ladder structures targeting the same transfer function. The last option was selected for prototype implementation, which will be described in Section~\ref{sec:practical_example}. Nominally ideal-element models were subjected to stray-resistance tests wherein measured commercial transformers were included and realistic stray capacitance added to certain critical nodes. The difference in performance is decisive in favour of the coupled-ladder structure exploiting the 1\nobreakdash-3 transmission path. The coupled-ladder is thus recommended for low-pass and band-pass designs. It is, however, true that the complexity of coupled-ladder realizations is higher, requiring, typically, several wideband transformers (or equivalently custom coupled inductors), but the difference in the resulting performance is deemed worthwhile.

Finally, it should be mentioned that the economy bridge structure can be completed to become a true biconjugate network, now fully 4-port equivalent to the coupled ladder structure. The cost is replacing the main transformer by a true hybrid coil, and using a second one to combine the outputs 3' and 4' of the reference filter arms (see the named ``canonical realization'' in Fig. 27 of \cite{belevitch1958}).

\section{Implementation Strategies for Elliptic/Quasi-Elliptic Functions}\label{sec:elliptic}

%The technique presented in this paper applies in general to transmission functions which have attenuation poles and/or reflection zeros at finite frequencies (added to those at the origin and at infinity). We will comment here on specific issues arising from cells that generate non-zero, finite frequency poles. Details for cell cases by 0 Hz or $\infty$ poles can be seen in [ ].
\begin{table*}[!t]
    \caption{Series Cell Adjunction}
    \label{tab:series}
    \centering
    \begin{tabular}{c|cccc}
        \hline\hline
        Ladder Increment & \multicolumn{4}{c}{Constant-Resistance Network Construction}\\
        \hline
        \includegraphics{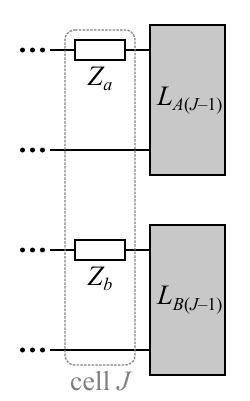} &
        \includegraphics{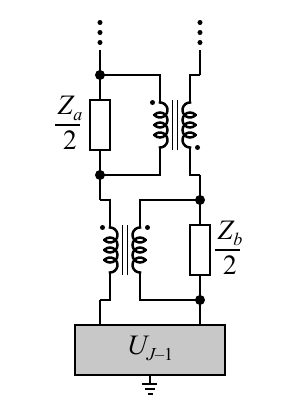} &
        \includegraphics{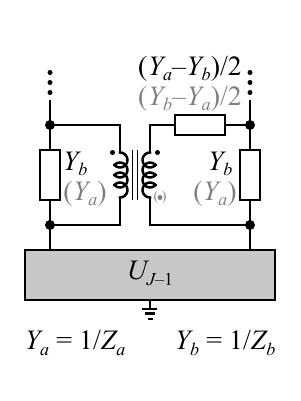} &
        \includegraphics{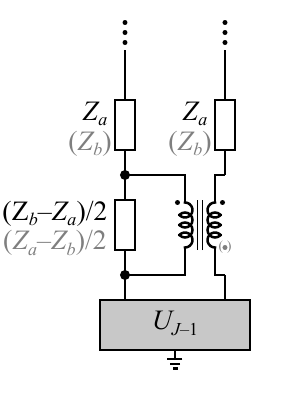} &
        \includegraphics{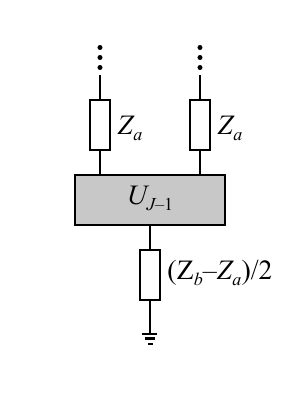}\\
        & Configuration 1$^\dagger$ & Configuration 2 & Configuration 3 & Configuration 4\\
        \hline
        Typical Cell Cases\\
        \cline{1-1}
        \raisebox{-0.3in}{\includegraphics{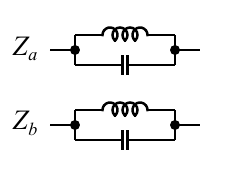}} &
        \begin{minipage}[c]{1.2in}\raggedright
            $\bullet$ Always realizable, even if pole frequencies are different\\
            $\bullet$ Can be realized with transformers, or coupled-coils having finite inductance\\
            $\bullet$ Good high-frequency isolation
        \end{minipage} &
        \begin{minipage}[c]{1.2in}\raggedright
            $\bullet$ Always realizable if poles have the same frequency, conditionally if the poles are different\\
            $\bullet$ Can be realized with transformers, or coupled-coils having finite inductance\\
            $\bullet$ Best high-frequency isolation
        \end{minipage} &
        \begin{minipage}[c]{1.2in}\raggedright
            NOT RECOMMENDED\\
            $\bullet$ Prone to parasitic resonance due to small mistuning.\\
            $\bullet$ Realizable only if poles have the same frequency
        \end{minipage} &
        \begin{minipage}[c]{1.2in}\raggedright
            NOT RECOMMENDED\\
            $\bullet$ Prone to parasitic resonance due to small mistuning.\\
            $\bullet$ Loss of common ground reference\\
            $\bullet$ Realizable only if poles have the same frequency
        \end{minipage}\\
        \\
        \hline
        \raisebox{-0.45in}{\includegraphics{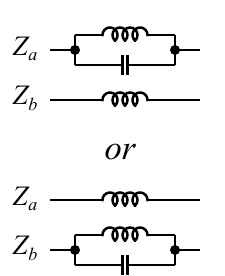}} &
        \begin{minipage}[c]{1.2in}\raggedright
            $\bullet$ Always realizable\\
            $\bullet$ Can be realized with transformers, or coupled-coils having finite inductance\\
            $\bullet$ Lone inductor's stray capacitance may limit high-frequency isolation.
        \end{minipage} &
        \begin{minipage}[c]{1.2in}\raggedright
            $\bullet$ Realizable if $L_{tank}<L_{single}$\\
            $\bullet$ Lone inductor's stray capacitance may limit high-frequency isolation.
        \end{minipage} &
        \begin{minipage}[c]{1.2in}\raggedright
            NOT REALIZABLE except when single inductor has degenerated to a short.
        \end{minipage} &
        \begin{minipage}[c]{1.2in}\raggedright
            NOT REALIZABLE except when $Z_a$ degenerates to a short and $Z_b$ is a tank.
        \end{minipage}\\
        \\
        \hline\hline
    \end{tabular}\\
    $^\dagger$Additional well-known equivalents for configuration 1 can be found in \cite{cauer}.
\end{table*}
\begin{table*}[!t]
    \caption{Shunt Cell Adjunction}
    \label{tab:shunt}
    \centering
    \begin{tabular}{c|ccc}
        \hline\hline
        Ladder Increment & \multicolumn{3}{c}{Constant-Resistance Network Construction}\\
        \hline
        \includegraphics{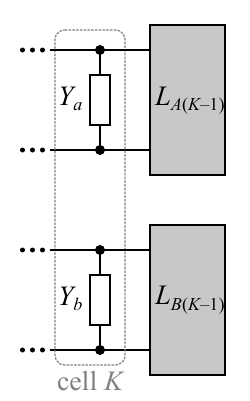} &
        \includegraphics{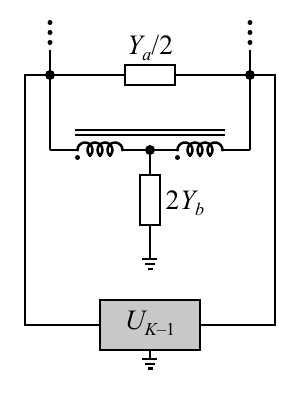} &
        \includegraphics{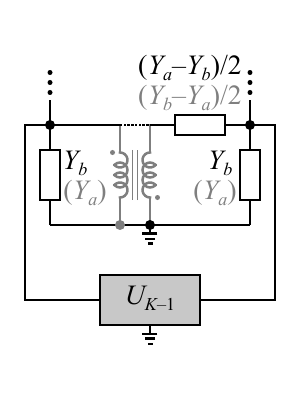} &
        \includegraphics{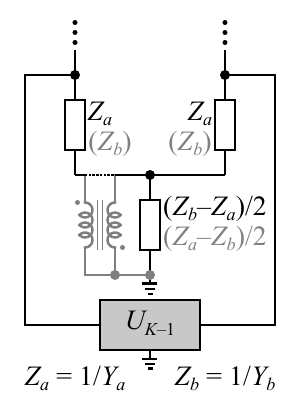}\\
        & Configuration 1$^\dagger$ & Configuration 2 & Configuration 3\\
        \hline
        Typical Cell Cases\\
        \cline{1-1}
        \raisebox{-0.3in}{\includegraphics{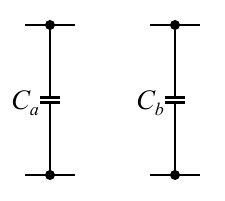}} &
        \begin{minipage}[c]{1.5in}\raggedright
            $\bullet$ Always realizable
        \end{minipage} &
        \begin{minipage}[c]{1.5in}\raggedright
            $\bullet$ Always realizable\\
            $\bullet$ Transformer necessary when $C_b>C_a$
        \end{minipage} &
        \begin{minipage}[c]{1.5in}\raggedright
            $\bullet$ Always realizable\\
            $\bullet$ Transformer necessary when $C_b>C_a$
        \end{minipage}\\
        \
        \raisebox{-0.3in}{\includegraphics{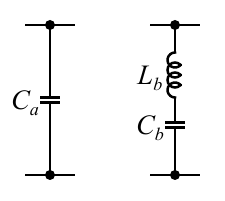}} &
        \begin{minipage}[c]{1.5in}\raggedright
            $\bullet$ Always realizable
        \end{minipage} &
        \begin{minipage}[c]{1.5in}\raggedright
            $\bullet$ NOT REALIZABLE
        \end{minipage} &
        \begin{minipage}[c]{1.5in}\raggedright
            $\bullet$ Realizable if $C_b<C_a$\\
            $\bullet$ Simple, free of coupling
        \end{minipage}\\
        \
        \raisebox{-0.3in}{\includegraphics{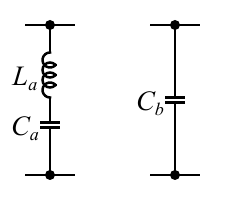}} &
        \begin{minipage}[c]{1.5in}\raggedright
            $\bullet$ Always realizable
        \end{minipage} &
        \begin{minipage}[c]{1.5in}\raggedright
            $\bullet$ NOT REALIZABLE
        \end{minipage} &
        \begin{minipage}[c]{1.5in}\raggedright
            $\bullet$ Realizable if $C_b>C_a$ and using transformer
        \end{minipage}\\
        \
        \raisebox{-0.3in}{\includegraphics{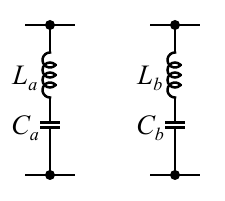}} &
        \begin{minipage}[c]{1.5in}\raggedright
            $\bullet$ Always realizable, even if pole frequencies are different
        \end{minipage} &
        \begin{minipage}[c]{1.5in}\raggedright
            NOT RECOMMENDED\\
            $\bullet$ Always realizable if pole frequencies are the same\\
            $\bullet$ Prone to parasitic resonance due to small mistuning\\
            $\bullet$ Transformer necessary when $L_b<L_a$
        \end{minipage} &
        \begin{minipage}[c]{1.5in}\raggedright
            $\bullet$ Always realizable if pole frequencies are the same, conditionally if they are different\\
            $\bullet$ No coupling if $L_b>L_a$ and pole frequencies are the same
        \end{minipage}\\
        \\
        \hline\hline
    \end{tabular}\\
    $^\dagger$Additional well-known equivalents for configuration 1 can be found in \cite{cauer}.
\end{table*}
In Tables~\ref{tab:series} and \ref{tab:shunt} we summarize the lattice equivalences needed for filter implementation, as each elementary lattice has to be converted into an equivalent two-port with the minimum number of components and adequate structure. Equivalences are given in a generic form, with incremental ladder immittances now labelled $Z_a$, $Y_a$, etc.
\begin{itemize}
    \item Elementary lattices connected in shunt (Table~\ref{tab:shunt}) will be converted to unbalanced form.
    \item Those connected in series (Table~\ref{tab:series}) will have to provide transformer isolation if they are connected on-top (configurations 1\nobreakdash-3, Table~\ref{tab:series}). Alternatively, the series connection can be implemented by the addition of tee-equivalent arms, as per configuration 4 (supposing $Z_b-Z_a$ is realizable); this is the only configuration in Table~\ref{tab:series} which can be implemented without transformer coupling.
\end{itemize}
Some typical specific cell cases are highlighted in these tables (focusing on finite frequency attenuation poles), with their strong and weak points. The following summary conclusions, restricted to these cases, can be articulated:
\renewcommand{\theenumi}{\alph{enumi}}
\begin{enumerate}
    \item Configuration 1 (using Cauer's bridges) is a general solution for both connection kinds, but implies coupling. It is suitable even in cases where the cell corresponds to two different finite-frequency poles. This configuration has some canonical equivalents \cite{cauer}, listed in Appendix D for reference. The version reflected in Table~\ref{tab:series} is frequently suitable for realization with coupling coils instead of an ideal transformer, while still providing adequate isolation.
    \item For series connection (Table~\ref{tab:series}), the only configuration without coupling is number 4, but it is not recommended for parallel resonators. Therefore, series connection of an elementary lattice will imply coupling in general. Configuration 2 is the appropriate non-canonical alternative to configuration 1.
    \item For shunt connection (Table~\ref{tab:shunt}), configuration 3 is the appropriate alternative to configuration 1.
    \item For the just mentioned non-canonical alternatives, and when both cell branches are finite-pole resonators, realizability is assured and the likelihood of coupling-free realization increases, if they correspond to the same pole frequency.
\end{enumerate}

Let us now consider the preparation of the reference dual filters prior to the process of cell assignment. Since only electrical duality (as opposed to topological duality) is required, and if both ladders are to be designed while targeting the minimum number of inductors, then, whenever possible, partial removal with a capacitor is preferred. This would imply that the same kind of resonator would result for the same pole in both dual structures (e.g. a parallel resonator for an upper stop-band pole and a series resonator for a lower stop-band pole).

This last condition is feasible in the case of an even degree quasi-elliptic (and generalized similar) low-pass filter, due to the availability of a double pole at infinity. In fact, for these transfer functions, a dual reference filter is obtained simply by reversing the input and output ports, since the conditions for antimetry are satisfied (assuming, as is is usually the case, that all reflection zeroes lie on the $j\omega$ axis). Additionally, if condition (d) above is attempted, the pole removal sequence can be appropriately altered in the dual with respect to the original, so that after reversing the ports, the same sequence of finite pole frequencies still results from input to output. In that case, a single cell can be associated with a finite-frequency pole. This is the strategy chosen in the prototype described in Section~\ref{sec:practical_example}. It nevertheless can be the case that negative component values appear in some desired pole sequence. If so, one should give up some constraint, for instance condition (d), and/or use configuration 1 whenever necessary.

\begin{figure}[!t]
    \centering
    \includegraphics{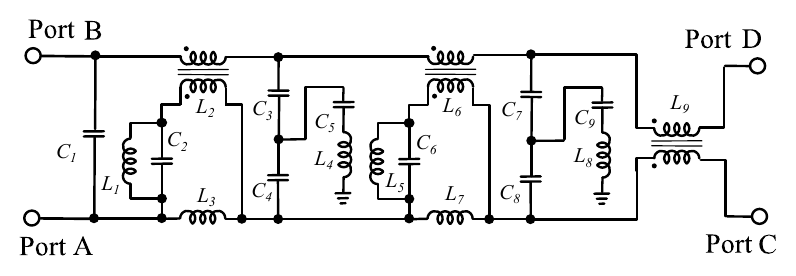}
    \caption{An odd degree reflectionless filter realization, tabulated in \cite{telefunkenkat} as ``C05 15 $\Theta $=55'' (values in Fig.~\ref{fig:refFildeg5}).}
    \label{fig:oddegfilter}
\end{figure}

In the case of an odd degree elliptic (generalized similar) low-pass filter, the poles must be realized in series branches as a parallel tank in one reference filter, and in shunt branches as a series resonator in the dual reference filter. A single cell cannot be associated with a specific pole, and the unrestricted usage of configuration 1 (or any preferred canonical equivalent) will likely be necessary at several places. Fig.~\ref{fig:oddegfilter}
shows the structure of a reflectionless realization of a fifth-degree elliptic low-pass filter tabulated in \cite{telefunkenkat} as ``C05 15 $\Theta$=55''. Some synthesis details are available in Appendix~\ref{app:odd_deg_data}.

If the odd degree low-pass response is not elliptic-like, and has several poles at infinity, then the strategy described for even degree is applicable.

In the same way the strategy can be extended to direct-band-pass filter designs, with two facts to be taken into account:
\begin{itemize}
    \item Band-pass reference designs usually require a higher degree, such that a higher number of coupling devices is needed.
    \item Direct-band-pass designs are also subject to unequal resistive terminations. Suitable Norton transformations have to be introduced so that both dual filters have the same terminations. These transformations increase the complexity and the number of required coupling devices in the derived reflectionless structure.
\end{itemize}
For these reasons, a preferable procedure could be to use band-pass designs derived via frequency transformation of a low-pass prototype. A practical method is to construct the low-pass reflectionless network first, and apply the frequency transformation to it directly. The number of required transformers is not affected by this last step. The resulting reflectionless band-pass structure maintains the same general complexity of the low-pass filter (degree halved), so in the end it is usually more economical, despite the increased number of inductors. Additionally, the issue with different resistive terminations is absent. Should resonances appear due to misalignments, applying the canonical configuration 1 to the concerned section already in the reflectionless low-pass prototype solves the issue.

Another possible strategy for band-pass designs with a very high relative bandwidth value could be to take advantage of the reflectionless cascading capability, and then to construct the band-pass filter by cascading a low-pass filter with a high-pass filter, each of them reflectionless. \cite{setty2018}.

Finally for band-stop designs, we believe that deriving them from a reflectionless high-pass design (via low-pass to band-pass frequency transformation) is a reasonable procedure in our context.

%Remarks :
%\begin{itemize}
%    \item Another realization, with the Cauer Lattice equivalent ``single Xfmr with secondary ratio +1/-1'' (which also offers isolation left-to-right), has not been included in the table, due to its bad practical behavior at high frequency.
%    \item Glitches are impairments presented by certain configurations which are more sensitive to unavoidable resonators misalignment.
%    \item Included in the second step case, we can consider when the single inductor reduces to a short. Then configuration 1 is the more adequate.
%    \item Preparation of reference filters
%    \begin{enumerate}
%        \item Cell splitting: recommended associations ?
%        \item Antimetry and topological duality.
%        \item Odd degree low-pass (non-antimetric) %case
%    \end{enumerate}
%    \item And for cases of non-coupling solutions searched:
%    \begin{enumerate}
%        \item Pole removal sequence adaptation
%        \item Forcing proportionality by partial finite pole removal, when pole removal sequence adaptation fails (to include or not ?): BETTER NOT.
%    \end{enumerate}
%    \item A Low-pass detailed design: LPRK6-A4 or A6?
%    \item Tolerances and Alignment
%    \item HP, BP and SB cases
%    \begin{enumerate}
%        \item HP : A-B path in an LP design ? or specific design A-C path?
%        \item BP: frequency transformation of LP to recommend  (strongly?)
%        \item SB: by means of a frequency transformation from an HP.
%    \end{enumerate}
%\end{itemize}

\section{Practical Implementation Example}\label{sec:practical_example}

\begin{figure}[tbp]
    \centering
    \includegraphics{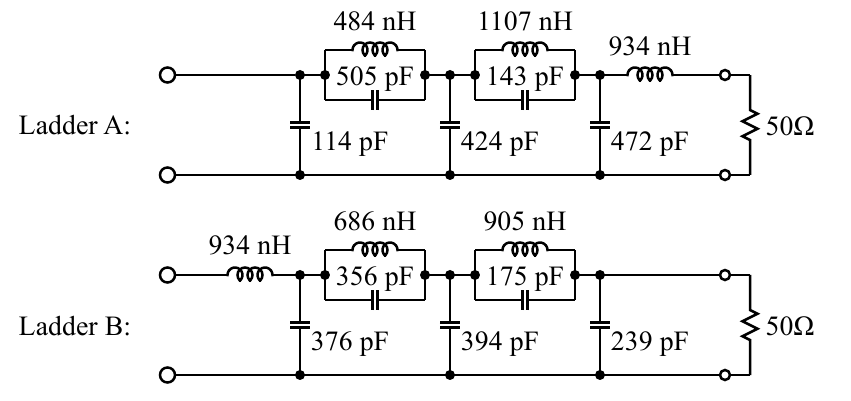}
    \caption{Reference ladders for the implemented sixth-order elliptic reflectionless filter.}
    \label{fig:test_ref_ladders}
\end{figure}

In order to test the theory described above, we have selected a sixth-order quasi-elliptic response for practical implementation. The reference ladders are shown in Fig.~\ref{fig:test_ref_ladders}.
Notably, the ladders have similar structures (floating parallel resonators in between grounded shunt capacitors) with an extra series element at the end of ladder A and the beginning of ladder B. These features facilitate the construction of elementary lattices having minimal complexity.

\begin{figure}[tbp]
    \centering
    \includegraphics{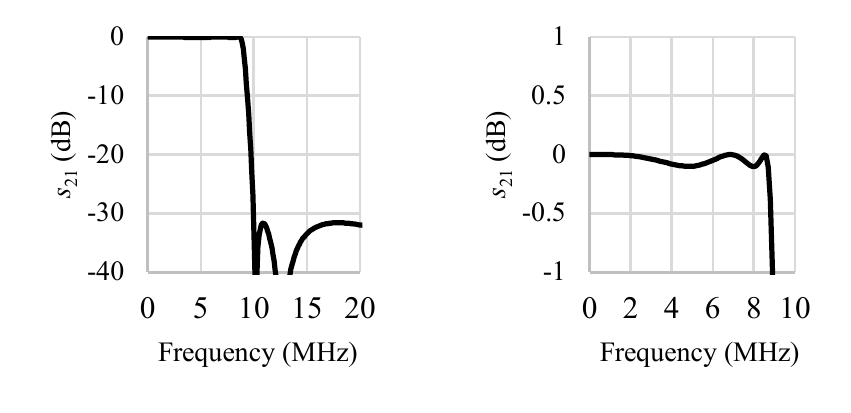}\\
    \hspace{0.4in}(a)\hspace{1.6in}(b)
    \caption{(a) Elliptic transmission response of the reference ladders for the implemented sixth-order reflectionless filter. The amplitude response of both ladders is the same. (b) Detail of pass-band.}
    \label{fig:test_ideal_response}
\end{figure}

The transmission response of these reference ladders is given in Fig.~\ref{fig:test_ideal_response}.
It has pass-band ripple of about 0.1 dB, while the stop-band ripple provides minimum rejection of about 31.6 dB, with steep cutoff near 10 MHz.

\begin{figure}[tbp]
    \centering
    \includegraphics{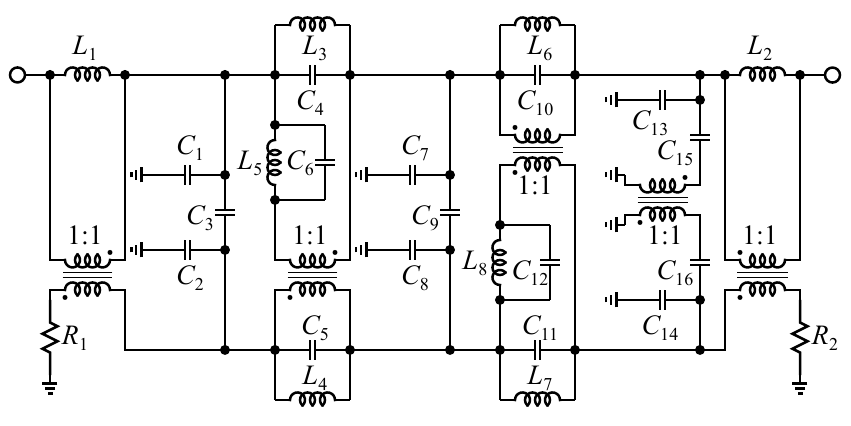}
    \caption{Coupled ladder reflectionless filter constructed from the reference ladders in Fig.~\ref{fig:test_ref_ladders}.}
    \label{fig:test_filter}
\end{figure}

The proposed coupled-ladder solution is shown in Fig.~\ref{fig:test_filter}.
and the final values for all components are given in Table~\ref{tab:test_elements}.
\begin{table}[!t]
    \caption{Final Element Values and Parts for the Sixth-Order Pseudo-Elliptic Reflectionless Filter}
    \label{tab:test_elements}
    \centering
    \begin{tabular}{ccc}
        \hline\hline
        Element & Ideal Value & Actual Value\\
        \hline
        $C_1$, $C_2$ & 239.4 pF & 240 pF\\
        $C_3$ & 116.1 pF & 120 pF\\
        $C_4$, $C_5$ & 142.7 pF & 150 pF\\
        $C_6$ & 15.94 pF & 16 pF\\
        $C_7$, $C_8$ & 394.0 pF & 390 pF\\
        $C_9$ & 15.11 pF & 15 pF\\
        $C_{10}$, $C_{11}$ & 355.9 pF & 360 pF\\
        $C_12$ & 74.45 pF & 75 pF\\
        $C_{13}$, $C_{14}$ & 114.0 pF & 110 pF\\
        $C_{15}$, $C_{16}$ & 262.3 pF & 270 pF\\
        $L_1$, $L_2$ & 466.8 nH & (tuned)\\
        $L_3$, $L_4$ & 1107 nH & (tuned)\\
        $L_5$ & 9909 nH & (tuned)\\
        $L_6$, $L_7$ & 685.7 nH & (tuned)\\
        $L_8$ & 3278 nH & (tuned)\\
        $R_1$, $R_2$ & $50\Omega$ & $49.9\Omega$\\
        \hline\hline
    \end{tabular}
\end{table}
All of the transformers have unity turns ratio, for which the authors selected part number TC1-6X+ from Mini-Circuits. Ceramic capacitors, size 0805, with 1\% tolerance were used, and the terminations were made with $49.9\Omega$ thin-film resistors, also 1\% tolerance.

\begin{figure}[tb]
    \centering
    \includegraphics{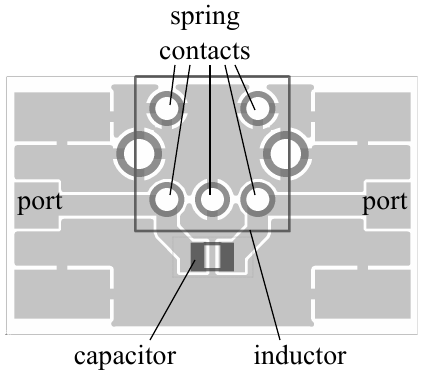}
    \includegraphics{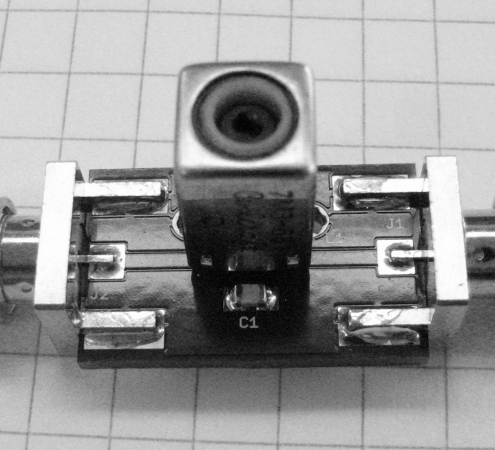}\\
    (a)\hspace{1.6in}(b)
    \caption{(a) Layout of inductor-tuning circuit board. (b) Photo with mounted inductor.}
    \label{fig:tuner}
\end{figure}

The selectivity of the transition demands elements of the highest precision, higher than is readily available from inexpensive, commercially available inductors, especially at these frequencies. We have therefore chosen to use shielded, tunable inductors instead (series 7M3 from Coilcraft). To facilitate tuning, a separate circuit board was fabricated comprising a parallel $LC$-resonator, shown in Fig.~\ref{fig:tuner}.
The inductor pin locations on the circuit board were equipped with tiny, press-in spring contacts (part no. 0531-0-15-15-31-27-10-0 from Mill-Max). Each inductor was then individually plugged into this board along with a tight-tolerance (1\%) reference capacitor, and tuned while monitoring the resonance on a Vector Network Analyzer (VNA). Once the resonance matched the predicted value of a given desired inductance, the inductor was withdrawn from the tuner board and permanently soldered into the final circuit. Multiple trials showed repeatability in the achieved resonance of $\pm10$ kHz, or typically $\pm0.05\%$, which is 20 times better than the reference capacitor itself or the other capacitors in the filter.

\begin{figure}[tb]
    \centering
    \includegraphics{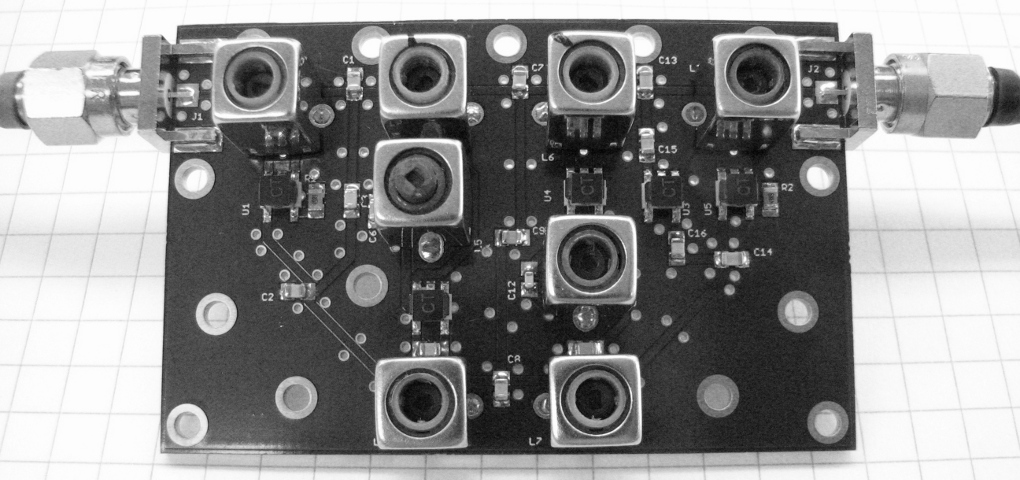}
    \caption{Photograph of completed pseudo-elliptic reflectionless filter circuit board.}
    \label{fig:test_filter_photo}
\end{figure}

\begin{figure}[tb]
    \centering
    \includegraphics{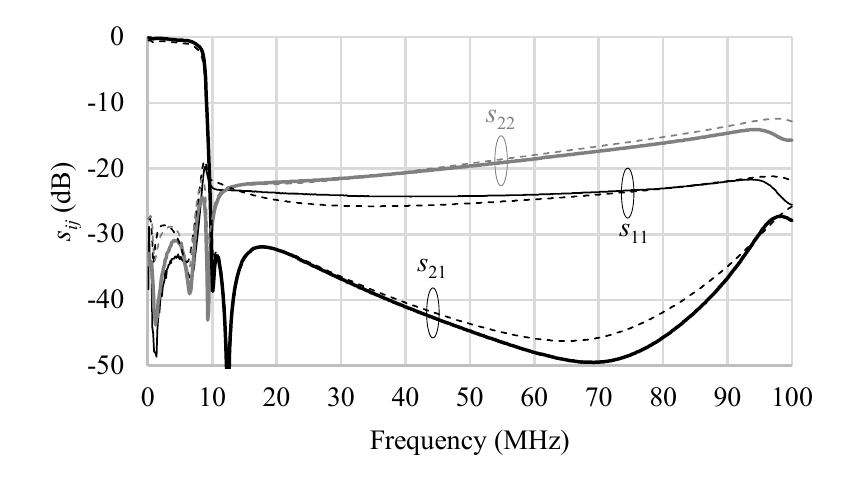}
    \caption{Measured (solid lines) and simulated (dashed lines) frequency response for the sixth-order pseudo-elliptic reflectionless filter.}
    \label{fig:test_meas}
\end{figure}

The final filter circuit board is shown in Fig.~\ref{fig:test_filter_photo}
and its measured response is in Fig.~\ref{fig:test_meas}.
Also plotted in dashed lines are the simulated results using detailed capacitor and inductor circuit models, which include their parasitics, and typical measured data for the transformers which are available from the vendor as a touchstone file. With the further addition of realistic electrical lengths for the traces on the circuit board, the agreement between measurement and theory throughout the pass-band, transition-band, and stop-band is excellent (in fact, the pass-band insertion loss is even less than expected). In particular, the nearly 20 dB return loss over the first two octaves beyond the cutoff into the stop-band is an exemplary demonstration of the critical feature which distinguishes this ``reflectionless'' topology from more conventional ones.

\begin{figure}[tb]
    \centering
    \includegraphics{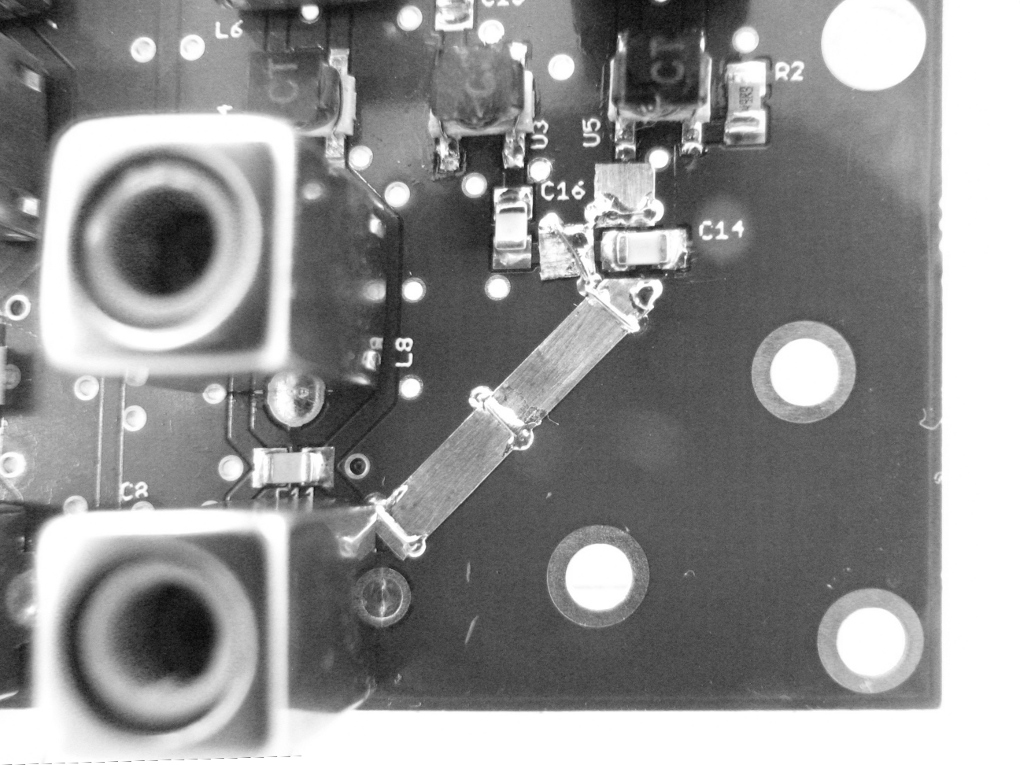}
    \caption{Detail of copper strips added to long traces on the circuit board in an attempt to reduce their parasitic inductance and stray coupling.}
    \label{fig:copper_detail}
\end{figure}

\begin{figure}[tb]
    \centering
    \includegraphics{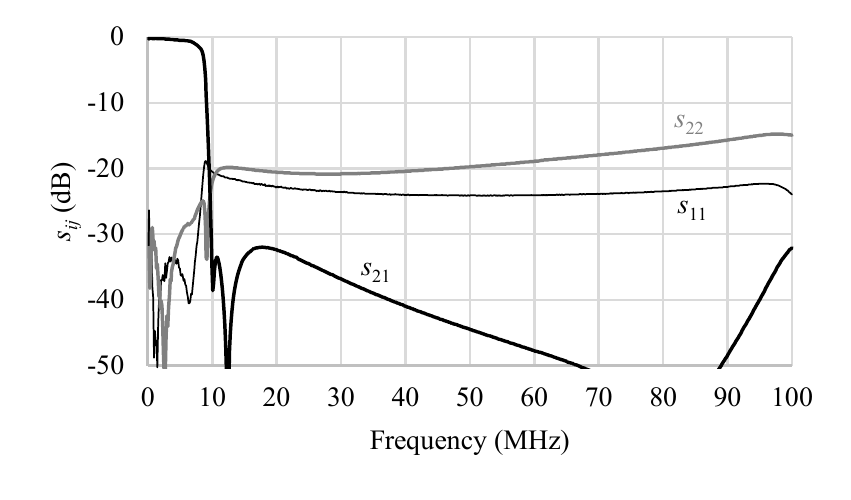}
    \caption{Reflectionless filer response after applying copper tape to the longest traces.}
    \label{fig:shifted}
\end{figure}

Like any low-pass filter, layout parasitics including trace inductance and inter-component coupling will ultimately lead to a degradation in stop-band performance at high frequency. An attempt was made to slightly mitigate these effects by applying strips of copper tape over the longest traces and then shorting them to the via fence in the top-side ground-plane (detail shown in Figure~\ref{fig:copper_detail}).
This had the effect of shifting the onset of stop-band degradation to higher frequencies, as indicated in Figure~\ref{fig:shifted}.
Thus, the stop-band rejection is better than 31 dB for a full decade beyond the cutoff frequency. If further stop-band extension was needed, the parasitic degradation could be further suppressed by cascading another small, lower-order reflectionless filter with higher cutoff.

\begin{figure}[tb]
    \centering
    \includegraphics{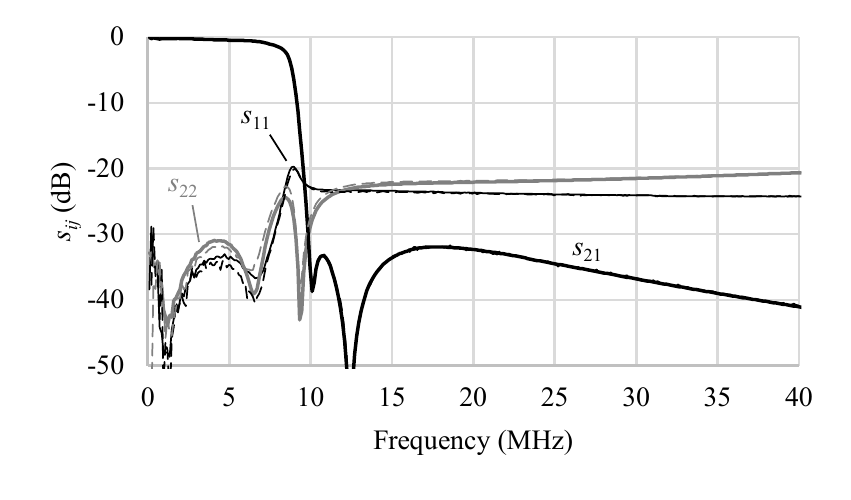}
    \caption{Comparison of results from two constructed filters (one in solid lines, the other dashed), showing excellent repeatability in both transmission and reflection.}
    \label{fig:test_twice}
\end{figure}

In order to further demonstrate the repeatability of the design, a second circuit board was assembled using the same layout and tuning technique (without the hand-cut copper tape shielding). A detail of the two responses is shown in Fig.~\ref{fig:test_twice}
highlighting that the manufacturing repeatability is excellent.

\section{Conclusion}

A new theorem for the development of coupled-ladder, reflectionless filter topologies has been described, and illustrated by examples. Tables of elementary lattices --- equivalent circuits which are building blocks of the coupled-ladder topology --- are presented to aid the reader in developing reflectionless filters to implement a wide variety of custom responses. A practical test circuit comprising a sixth-order, low-pass, reflectionless filter having a pseudo-elliptic frequency response has been constructed, and measurements show excellent agreement with the theory.

\appendices
\section{The Lattice Reduction Method, a Predecessor}\label{app:lattice_reduction}

The lattice reduction method has been known for a long time. L. Weinberg included it in is his doctoral thesis \cite{weinberg1951new} from 1951, as a systematic transformation algorithm. He applied it for the first time to derive an unbalanced equivalent of a lattice whose arms are ladder networks, by progressive construction (see Fig.24 of \cite{weinberg1951new}).  As presented, however, it requires proportionality between cell branches, with a real positive factor $Z_b/Z_a\ge1$, since only transformer-free realization was considered. Consequently,  only part of the possible equivalences needed for network construction were presented. The \emph{elementary lattice} breakdown presented in Section~\ref{sec:theorem} has a wider scope, without such restrictions, and more importantly points to the true objective of the transformation: a ladder of elementary symmetrical two-ports. Nevertheless, the rules of this predecessor are still of interest, in that, for example, they allow one to quickly derive some of the useful equivalents for typical elementary lattices. These rules are summarized in Table~\ref{tab:lattice_reduction},
\begin{table}[!t]
    \caption{Basic Equivalences for Lattice Reduction to Unbalanced Form}
    \label{tab:lattice_reduction}
    \centering
    \begin{tabular}{@{}c|c|c|c@{}}
        \hline\hline
        Case & Lattice Form & & Unbalanced Equivalent\\
        \hline
        0 & \raisebox{-0.4in}{\includegraphics{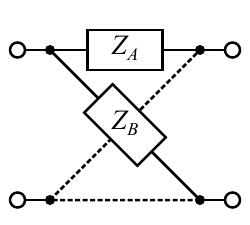}} & \makecell[c]{If\\\\$\leftrightarrow$\\\\Then:} & \raisebox{-0.2in}{\includegraphics{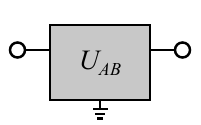}}\\
        \hline
        1 & \raisebox{-0.4in}{\includegraphics{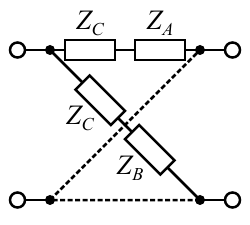}} & $\rightarrow$ & \raisebox{-0.2in}{\includegraphics{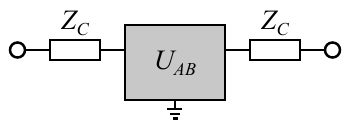}}\\
        \hline
        2 & \raisebox{-0.4in}{\includegraphics{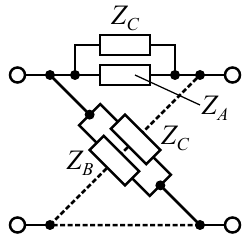}} & $\rightarrow$ & \raisebox{-0.2in}{\includegraphics{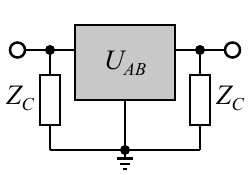}}\\
        \hline
        3 & \raisebox{-0.4in}{\includegraphics{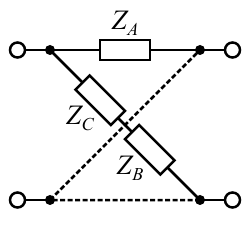}} & $\rightarrow$ & \raisebox{-0.3in}{\includegraphics{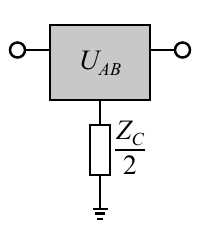}}\\
        \hline
        4 & \raisebox{-0.4in}{\includegraphics{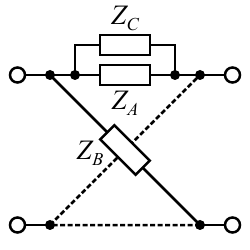}} & $\rightarrow$ & \raisebox{-0.2in}{\includegraphics{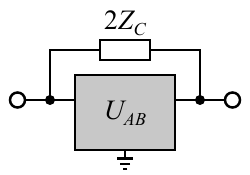}}\\
        \hline
        5 & \raisebox{-0.4in}{\includegraphics{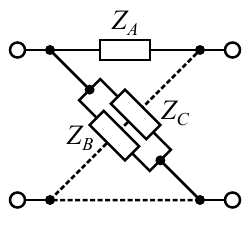}} & $\rightarrow$ & \raisebox{-0.3in}{\includegraphics{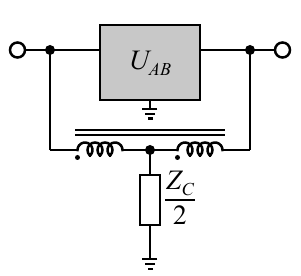}}\\
        \hline
        6 & \raisebox{-0.4in}{\includegraphics{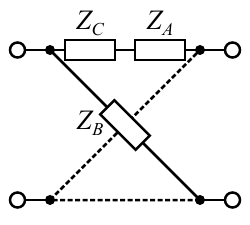}} & $\rightarrow$ & \raisebox{-0.3in}{\includegraphics{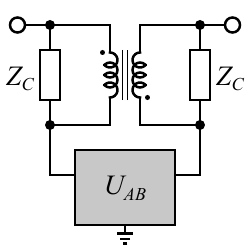}}\\
        \hline\hline
    \end{tabular}
\end{table}
along with two rather natural additions (cases 5 and 6) solving realization when $Z_b/Z_a<1$ (but still real positive), and which were not articulated in the past.

\section{Demonstration of Equivalence Theorem}\label{app:demonstration}

Demonstration of the equivalence theorem from Section~\ref{sec:equivalence_theorem} is constructive, and proceeds by mathematical induction. Let us assume we know an unbalanced equivalent of the lattice network whose arms are that portion of ladders A and B from the terminations through cell $(j-1)$. Let us further designate $Z_{A(j-1)}$ and $Z_{B(j-1)}$ the input impedance of these partial ladders. Then, when a new series cell is added to the partial arm ladders, we have
\begin{subequations}
    \begin{equation}
        Z_{Aj} = Z_{A(j-1)}+\Delta Z_{Aj}
    \end{equation}
    \begin{equation}
        Z_{Bj} = Z_{B(j-1)}+\Delta Z_{Bj}
    \end{equation}
\end{subequations}
The $Z$-parameter matrix of any equivalent to the lattice with augmented ladder arms may then be decomposed in the following way,
\begin{multline}
    \frac{1}{2}\begin{pmatrix}
        Z_{Bj}+Z_{Aj} & Z_{Bj}-Z_{Aj}\\
        Z_{Bj}-Z_{Aj} & Z_{Bj}+Z_{Aj}
    \end{pmatrix}\\ = \frac{1}{2}\begin{pmatrix}
        Z_{B(j-1)}+Z_{A(j-1)} & Z_{B(j-1)}-Z_{A(j-1)}\\
        Z_{B(j-1)}-Z_{A(j-1)} & Z_{B(j-1)}+Z_{A(j-1)}
    \end{pmatrix}\\
    +\frac{1}{2}\begin{pmatrix}
        \Delta Z_{Bj}+\Delta Z_{Aj} & \Delta Z_{Bj}-\Delta Z_{Aj}\\
        \Delta Z_{Bj}-\Delta Z_{Aj} & \Delta Z_{Bj}+\Delta Z_{Aj}
    \end{pmatrix}
\end{multline}
The first term corresponds to the $Z$-parameters of the network before augmentation, and the second one to an elementary lattice with arms $\Delta Z_{Aj}$ and $\Delta Z_{Bj}$. The incremental part is obviously a realizable two-port, and the matrix sum indicates that the augmented network can then be built by connecting these two two-ports in series.

In the same way, and operating with $Y$-parameters instead of $Z$-parameters, the case of a shunt-cell adjunction is solved, leading to the parallel connection of the elementary lattice with the original lattice. Applying this construction algorithm repetitively from the initial ``resistive termination lattice'' up to the complete ladders provides the desired equivalent global network, expressed by the development of \eqref{eq:NW_zparams} and \eqref{eq:NW_yparams} in Section~\ref{sec:equivalence_theorem}.

\section{Synthesis data for an odd degree elliptic reflectionless realization}\label{app:odd_deg_data}

\begin{figure}[!t]
        \centering
        \includegraphics{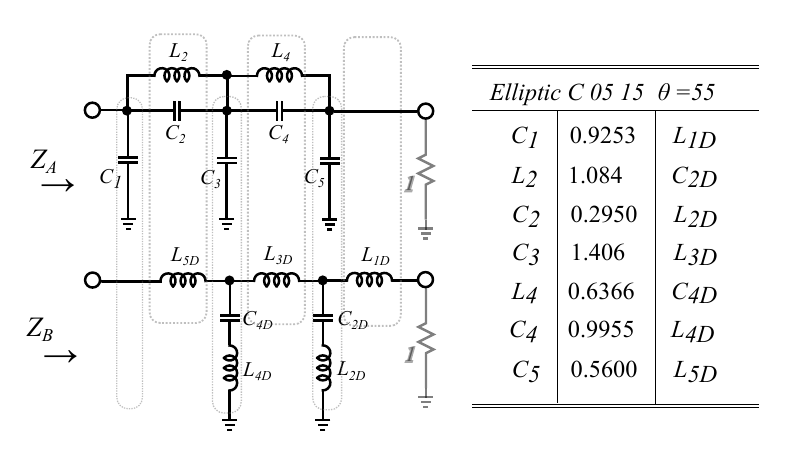}
        \caption{Reference filters for the topology in Fig.~\ref{fig:oddegfilter} (annotated C05 15 $\Theta$=55).}
        \label{fig:refFildeg5}
    \end{figure}

The two dual reference filters, taken from tables in \cite{telefunkenkat}, are reflected in Fig.~\ref{fig:refFildeg5}.
The dual filter $Z_B$ has been reversed between input/output ports, to get the minimum number of coupling devices in the reflectionless development. This is possible because both references are symmetrical filters, having odd degree. This is similar to the pole removal sequence adaptation, mentioned previously in Section~\ref{sec:elliptic}.

With the cell assignment shown in Fig.~\ref{fig:refFildeg5}, the reflectionless filter of Fig.~\ref{fig:oddegfilter} results. The denormalized element values are listed in Table~\ref{tab:c0515theta55}.
\begin{table}[!t]
    \caption{Elements of the Reflectionless Filter from Fig.~\ref{fig:oddegfilter}, Denormalized to $f_0=8.200$ MHz and $R_0=50\Omega$}
    \label{tab:c0515theta55}
    \centering
    \begin{tabular}{cc|cc}
        \hline\hline
        Element & Value ($\mu$H) & Element & Value (pF)\\
        \hline
        $L_1$ & 3.9601 & $C_1$ & 108.7\\ 
        $L_2$ & 0.89796 & $C_2$ & 193.2\\ 
        $L_3$ & 0.89796 & $C_3$ & 545.8\\ 
        $L_4$ & 0.14314 & $C_4$ & 545.8\\ 
        $L_5$ & 9.1868 & $C_5$ & 3675\\ 
        $L_6$ & 1.3644 & $C_6$ & 57.26\\  
        $L_7$ & 1.3644 & $C_7$ & 359.2\\  
        $L_8$ & 0.48304 & $C_8$ & 359.2\\ 
        $L_9$ & 0.2717 & $C_9$ & 1584\\  
        \hline\hline
    \end{tabular}
\end{table}
The turns-ratio of all coupled coils is 1:1. The filter response is similar to that of the prototype sixth-degree filter described in Section~\ref{sec:practical_example}, with only a slightly less-steep cut-off transition.

%\section{Cauer’s Equivalents for a Symmetric Lattice}\label{app:cauer_equivalents}

%For completeness, we wish to acknowledge and remind the reader of additional canonical equivalents to a symmetric lattice which are due to Cauer \cite{cauer} for configuration 1 from tables ~\ref{tab:series} and ~\ref{tab:shunt}. These are reflected in Fig.~\ref{fig:CauerEquiv}. We note that Figs.~\ref{fig:CauerEquiv}(a) and (b) cannot be realized by means of finite coupled coils and do therefore require a more ideal transformer in general. 
%\begin{figure}[!t]
%        \centering
%        \includegraphics{fig16}
%        \caption{Cauer Equivalents to canonic configuration 1 in Table~\ref{tab:series} : (a), and Table~\ref{tab:shunt} : (b,c)}
%        \label{fig:CauerEquiv}
%\end{figure}

\section*{Acknowledgment}

The primary author would like to express his grateful remembrance of Dr. E. Christian, once professor of Electrical Engineering at North-Carolina State University, for his teaching style and enduring influence.

%\IEEEtriggeratref{14}

\bibliographystyle{IEEEtran}
%\bibliography{morgan,other}
%=======================================
% Pasted from output.bbl

%=======================================
\begin{IEEEbiography}[{\includegraphics[width=1in,height=1.25in,clip,keepaspectratio]{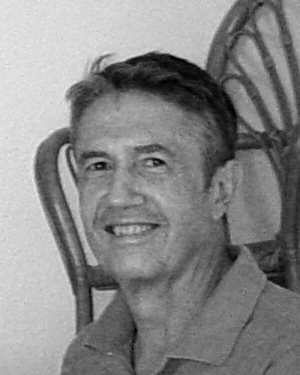}}]{Augusto Guilabert} received his M.S. in Telecom Engineering from the E.T.S.I.T (High School for Telecom Engineering), U.P.M. Madrid in 1969. He began to work for Standard El\'ectrica S.A., a subsidiary in Spain of ITT Corp., as a development engineer for land transmission systems. From 1973 to 1976 he worked in the German ITT subsidiary, SEL Stuttgart, on high capacity analog transmission systems. There, between other projects, he was nominated to develop the Quaternary Through-Group Filter, a hard bone for filter design at that time. After launching production, he returned to Standard El\'ectrica in Madrid, to participate in the development of most active and passive functions of land transmission equipment. Following the decline of analog telephony developments by 1979, he changed to mobile radio, where he was nominated to lead the development of the future ITT PMR (train-to-ground) system for the Spanish National Railways Network. There, he was the chief-development engineer, for two generations of equipment, throughout the 1980's. This system is still being utilized in the spanish conventional railway lines. By 1992 he had begun working in digital cordless telephony,  and by 1996 in wireless local loop systems, alternating system design and project leading functions. In 2001 he left the company which had belonged since the latter 1980's to Alcatel N.V., and joined 3Bymesa S.A., an independent smaller company working in the field of industrial electronics. There, he worked in manufacturing automation and telemetry applications. He developed sensor and data logging modules for road traffic monitoring, for an international market supplier. He retired in 2012, but remains active in the captivating field of his origins, network theory.
\end{IEEEbiography}

\begin{IEEEbiography}[{\includegraphics[width=1in,height=1.25in,clip,keepaspectratio]{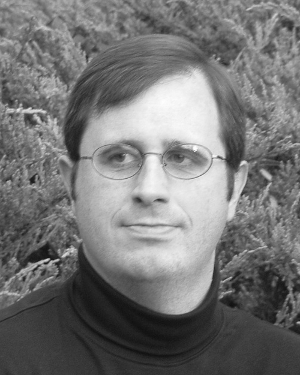}}]{Matthew A. Morgan}
(M'99--SM'17) received his B.S. in electrical engineering from the University of Virginia in 1999, and his M.S. and Ph.D. from the California Institute of Technology in 2001 and 2003, respectively.

During the summers of 1996 through 1998, he worked for Lockheed Martin Federal Systems in Manassas, VA, as an Associate Programmer, where he wrote code for acoustic signal processing, mathematical modeling, data simulation, and system performance monitoring. In 1999, he became an affiliate of NASA’s Jet Propulsion Laboratory in Pasadena, CA. There, he conducted research in the development of Monolithic Millimeter-wave Integrated Circuits (MMICs) and MMIC-based receiver components for atmospheric radiometers, laboratory instrumentation, and the deep-space communication network. In 2003, he joined the Central Development Lab (CDL) of the National Radio Astronomy Observatory (NRAO) in Charlottesville, VA, where he now holds the position of Scientist/Research Engineer. He is currently the head of the CDL’s Integrated Receiver Development program, and is involved in the design and development of low-noise receivers, components, and novel concepts for radio astronomy instrumentation in the cm-wave, mm-wave, and submm-wave frequency ranges. He has authored over 60 papers and holds seven patents in the areas of MMIC design, millimeter-wave system integration, and high-frequency packaging techniques. He is the author of \emph{Reflectionless Filters} (Norwood, MA: Artech House, 2017).

Dr. Morgan is a member of the International Union of Radio Science (URSI), Commission J: Radio Astronomy. He received a Topic Editor's Special Mention in the IEEE THz Transactions Best Paper competition and the Harold A. Wheeler Applications Paper Award in 2015.
\end{IEEEbiography}
\vfill

\begin{IEEEbiography}[{\includegraphics[width=1in,height=1.25in,clip,keepaspectratio]{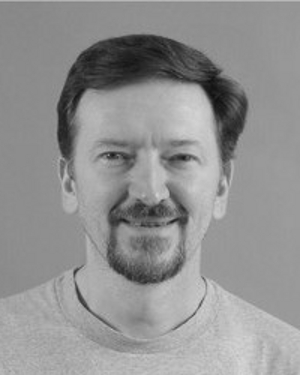}}]{Tod A. Boyd} was born in Steubenville, Ohio, in 1962. He received the A.S.E.E. degree from the Electronic Technology Institute, Cleveland, Ohio, in 1983.  From 1983 to 1985, he was with Hostel Electronics in Steubenville, Ohio.  In 1985, he joined Northrop Corporation’s Electronic Countermeasures Division, in Buffalo Grove, Ill., specialized in supporting the B-1B Lancer (secret clearance.)  In 1990, he joined Interferometrics, Inc., Vienna, Va., where he constructed VLBA tape recorders for the international Radio Astronomy community.

Since 1996, he has been with the National Radio Astronomy Observatory's Central Development Lab, Charlottesville, VA, where he initially assisted with the construction of cooled InP HFET amplifiers for the NASA's Wilkinson Microwave Anisotropy Probe (WMAP) mission.  Presently as a Technical Specialist IV he provides technical support for the advanced receiver R\&D initiatives. His responsibilities also include constructing low noise amplifiers for the Enhanced VLA and the Atacama Large Millimeter/sub-millimeter Array (ALMA) projects.
\end{IEEEbiography}
\vfill

%\newpage


%\vfill


\end{document}